\documentclass[sigconf]{acmart}
\usepackage{mathtools}
\usepackage{subcaption}
\usepackage{float}
\usepackage{stfloats}
\usepackage{makecell}
\usepackage{algorithm}
\usepackage{algpseudocode}
\usepackage{multirow}
\usepackage[utf8]{inputenc}
\usepackage{enumitem}

\AtBeginDocument{%
  }

\setcopyright{acmlicensed}
\copyrightyear{2025}
\acmYear{2025}
\acmDOI{XXXXXXX.XXXXXXX}
\acmConference[Conference acronym 'XX]{Make sure to enter the correct
  conference title from your rights confirmation email}{June 03--05,
  2025}{Woodstock, NY}
\acmISBN{978-1-4503-XXXX-X/2025/06}

\begin{document}

\title{Compute Only Once: UG-Separated TokenMixer for \\Efficient Large Recommendation Models}

\author{
  Hui Lu\textsuperscript{1*}, 
  Zheng Chai\textsuperscript{1*}, 
  Shipeng Bai\textsuperscript{1*}, 
  Hao Zhang\textsuperscript{2*}, 
  Zhifang Fan\textsuperscript{2}, 
  Kunmin Bai\textsuperscript{1},
  Ke Sun\textsuperscript{1},\\
  Yingwen Wu\textsuperscript{2}, 
  Bingzheng Wei\textsuperscript{2}, 
  Xiang Sun\textsuperscript{2}, 
  Ziyan Gong\textsuperscript{1}, 
  Tianyi Liu\textsuperscript{1}, 
  Hua Chen\textsuperscript{1}, \\
  Deping Xie\textsuperscript{1}, 
  Zhongkai Chen\textsuperscript{2}, 
  Zhiliang Guo\textsuperscript{2}, 
  Qiwei Chen\textsuperscript{2}, 
  Yuchao Zheng\textsuperscript{1†}
}

\thanks{*These authors contributed equally. \textsuperscript{†}Corresponding Author.}

\affiliation{%
  \institution{\textsuperscript{1}ByteDance AML \quad \textsuperscript{2}ByteDance}
  \country{}
}

\email{{luhui.xx, chaizheng.cz, baishipeng, zhanghao.229, zhengyuchao.yc}@bytedance.com}

\renewcommand{\shortauthors}{xx et al.}

\begin{abstract}
Driven by scaling laws, recommender systems increasingly rely on larger-scale models to capture complex feature interactions and user behaviors, but this trend also leads to prohibitive training and inference costs. While long-sequence models (e.g., LONGER) can reuse user-side computation through KV Caching, such reuse is difficult in TokenMixer-based dense feature interaction architectures (e.g., RankMixer/TokenMixer-Large), where user and group (i.e., candidate item) features are deeply entangled and mixed-up across layers. In this work, we present \textbf{User–Group Separation (UG-Sep)}, an industrial large-scale framework that enables user-side computation reusable in TokenMixer-based dense interaction models for the first time. UG-Sep explicitly disentangles user-side and item-side information flows within token-mixing layers, ensuring that a subset of tokens preserves purely user-side representations across layers. This design allows the corresponding per-token computations to be reused across multiple samples, significantly reducing redundant inference cost. To compensate for the potential expressive capacity loss induced by masking, we further propose an Information Compensation strategy that adaptively reconstructs suppressed user–item interactions. Moreover, as UG-Sep substantially reduces user-side FLOPs and exposes memory-bound components, we incorporate W8A16 (8-bit weight, 16-bit activation) weight-only quantization to alleviate memory bandwidth bottlenecks and achieve additional acceleration. We conduct extensive offline evaluations and large-scale online A/B experiments at ByteDance to validate the effectiveness of UG-Sep. Results show that UG-Sep reduces inference latency by up to 20\% without causing adverse changes to online user experience and commercial metrics on multiple influential business scenarios compared to TokenMixer at ByteDance, including 
Douyin Feed Recommendation, Hongguo Feed Recommendation, Chuanshanjia Ads, and Qianchuan Ads.

\end{abstract}





\maketitle

\section{Introduction}

Propelled by the scaling law, recommender system have continuously expanded in terms of parameter scale, training data volume, and computational cost, with widespread deployment in core processes such as retrieval \cite{lian2020lightrec,diaz2025recall,li2024harnessing,altosaar2021rankfromsets}, ranking \cite{jiang2026tokenmixer, zhu2025rankmixer}, and multitask learning \cite{wang2023towards,zhang2025optimizing}  in recent years. Such larger models can capture more complex feature interaction, user behavior patterns, and long-term preferences, significantly improving recommendation performance.
However, this growth has led to an exponential increase in training and inference costs, posing a critical challenge for model deployment.

Currently, methods for scaling up recommender models can be divided into two main directions: long sequence modeling \cite{chai2025longer,deng2025onerec,xu2025climber,chai2022user} and large dense feature interaction \cite{zhu2025rankmixer,zhang2024wukong}. In long sequence modeling, samples generated by the same user at different time points typically share substantial portions of their historical behavior sequences. Motivated by this feature, user-level sample aggregation \cite{zhai2024actions, chai2025longer} is widely employed during training, wherein samples belonging to the same user within a specified temporal window are aggregated and computed jointly. This aggregation enables computation reuse over shared sequence prefixes, leading to a substantial improvement in overall training efficiency. During the inference phase, KV-Cache mechanisms (e.g., M-Falcon \cite{zhai2024actions}) are used to cache the representations of the user-side historical sequences. This allows for the reuse of prefix computation results when processing each new candidate item, significantly reducing inference costs and latency.

However, such aggregation strategies are difficult to apply in large-scale dense feature interaction scenarios. TokenMixer modules like RankMixer\cite{zhu2025rankmixer}, TokenMixer-Large\cite{jiang2026tokenmixer}, and MLP-Mixer\cite{li2025research,gong2025pyramid,qi2025mtmixatt}, which involve deep cross-interaction between user-side and candidate-side features, perform intricate feature fusion at each layer through mechanisms such as channel mixing or nonlinear interactions. Thus, user-side features cannot be treated as independent prefixes that can be cached and reused. Whenever the candidate item changes, all cross-layer fusion computations must be recomputed, leading to a significant item dependency in the inference process. As a result, with the continuous increase in model width and depth, the inference overhead in dense scaling rises rapidly with model size, becoming a critical bottleneck for scaling up large-scale recommender systems.


To address the limitations, we introduce \textbf{UG-Sep}, the first technique designed to enable \textbf{User–Group}\footnote{A group refers to a deduplicated item, where each candidate item corresponds to a unique group\_id. In this paper, the terms "group" and "item" are used interchangeably.} \textbf{Separation}, and enable the reuse of user-side representations within TokenMixer architectures, which substantially reduces redundant computations and inference latency in large-scale dense feature interaction scenarios. Specifically, we \textbf{first} introduce a novel disentanglement mechanism that explicitly separate the information flows of the user side and item side within the model. This mechanism guarantees that a designated subset of tokens retains purely user-side information even after passing through arbitrary layer transformations, thereby enabling their associated Pertoken FFN \cite{zhu2025rankmixer} computations to be shared across multiple samples via user-level aggregation. 
\textbf{Second}, since the imposed masking inevitably leads to partial information loss and may diminish the model’s expressive capacity, we propose an Information Compensation strategy that adaptively reconstructs the suppressed or missing interactions between user and candidate features, ensuring that the model preserves robust predictive performance while benefiting from substantial computational savings.
\textbf{Last but not least}, with the adoption of UG-Sep, the computation FLOPs of the U-side are substantially reduced, causing certain components in TokenMixer to become GPU Memory-bound. To mitigate this emerging memory bottleneck, we incorporate the W8A16 (8-bit weight, 16-bit activation) quantization \cite{kim2025zero,yashaswini2025non} technique, which effectively alleviates GPU memory access cost and enables additional acceleration. Extensive offline and online experiments have validated the effectiveness of our approach. For example, in the production deployment of Douyin's feed recommendation system, UG-Sep has reduced inference latency by 20\% while maintaining stable online metrics. This demonstrates the feasibility and stability of the method in real-world production environments. In summary, the contributions are as follows:
\begin{itemize}
    \item We introduce UG-Sep, the first technique specifically designed for separable feature interaction in TokenMixer-based large recommendation models. Through a novel design with elaborate separation and reusable pertoken FFN mechanism, UG-Sep isolates user–item information flows and enables effective reuse of user-side computation that were previously inseparable in dense models.
    \item We further devise an information compensation strategy and introduce the W8A16 quantization technique, which synergistically elevate both the effectiveness and inference efficiency of UG-Sep to a higher level.
    \item The core UG-Sep mechanism is highly general and can be readily integrated into a wide range of interaction architectures, including TokenMixer variants, Transformer-based structures, etc., without requiring modifications to their fundamental design principles.
    \item Extensive empirical studies conducted both on large-scale industrial datasets and through online A/B testing on multiple high-traffic platforms at ByteDance demonstrate the practical value of the method.
\end{itemize}
\section{Related Work}


\subsection{Long-Sequence Modeling}
Modeling long-term user behavior sequences has been a central topic in recommender systems. Early approaches employ RNN and CNN-based architectures \cite{zhou2018deep, zhou2019deep, chang2021sequential, zhou2019deep,zhou2020can} to capture sequential dependencies. Besides, multi-domain \cite{chai2025adaptive,chang2023pepnet}, multi-interest \cite{li2019multi,chai2022user}, and sequence denoising methods \cite{shin2024attentive,chen2022denoising} are extensively approached for different aspects in modeling user preferences. As recommender systems continue to scale, several studies investigate the scaling behavior of sequential models \cite{zhai2024actions, zivic2024scaling}. To support industrial deployment, system-oriented solutions such as LONGER \cite{chai2025longer} focus on scaling up long-sequence modeling in production environments by optimizing data pipelines and training strategies, enabling practical use of ultra-long user histories.

However, the introduction of ultra-long user behavior sequences also brings substantial computational challenges. To address this, recent studies have explored solutions from the perspective of computation reuse and system-level optimization. For example, HSTU \cite{zhai2024actions} leverages user-level sequences to achieve efficient training and inference, thereby highlighting the importance of sequence-level computation reuse in extreme-scale generative recommendation settings. LONGER \cite{chai2025longer} explicitly targets the efficiency challenges of long-sequence modeling by adopting global tokens and system optimizations, enabling scalable training with ultra-long user histories. UniRec \cite{xie2024unifiedssr} proposes a unified framework for sequential recommendation that reuses user behavior sequences across different recommendation tasks and samples, thereby substantially improving training and inference efficiency. Overall, these studies consistently demonstrate that user-level sequence–centric organization and prefix reuse mechanisms are key factors in accelerating sequential recommendation models, laying a solid foundation for efficient scaling in long-sequence modeling.

\begin{figure*}[t]
    \centering
        \centering
        \includegraphics[width=1\linewidth ]{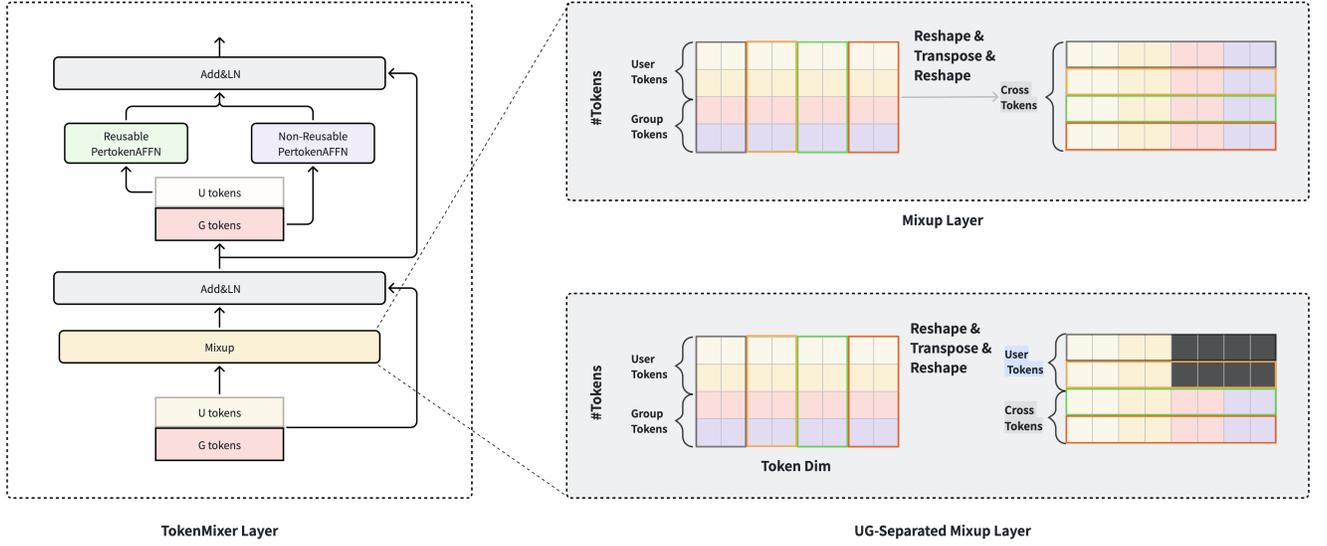}
        \caption{TokenMixer-Style Layer with UG-Sep}\label{fig:mixup}
\end{figure*}
\subsection{Dense Feature Interaction}

Beyond sequence-based approaches,  TokenMixer–style interaction models~\cite{tolstikhin2021mlp,zhu2025rankmixer} have gained increasing attention due to their strong expressive power. Such models perform deep cross-interaction between user-side and item-side features at every block, enabling fine-grained feature fusion. 


Despite their effectiveness, dense interaction models exhibit strong item dependency: user-side representations are repeatedly fused with candidate-side features throughout the network. As a result, user representations cannot be isolated as reusable prefixes. Whenever the candidate item changes, all cross-layer fusion computations must be recomputed, causing inference cost to scale linearly with the number of candidates. As model width and depth continue to increase, this property leads to rapidly growing inference overhead, becoming a major bottleneck for dense scaling in industrial recommender systems. To the best of our knowledge, existing work has not provided a general mechanism to enable user-side computation reuse in TokenMixer-based large scale recommendation architectures.

\section{Method}
\subsection{Separatation of UG Token}
In existing TokenMixer-based feature-interaction modules (\cite{zhu2025rankmixer,jiang2026tokenmixer,qi2025mtmixatt,zhang2025billion}), the input tokens typically encode a mixture of both U-side and G-side information. However, UG-Sep requires that the inputs to these interaction modules be cleanly separated into user-side and group-side representations. To satisfy this requirement, we first decouple the U-side and G-side features at the bottom layers of the network. Correspondingly, the feature-extraction modules (e.g., SENet \cite{chen2025squeeze}, DCN \cite{wang2017deep}) are split into two branches: one dedicated to extracting U-side features and the other to extracting G-side features. Also, for modules whose structural complexity prevents a clean separation between U-tokens and G-tokens, we treat their outputs as G-side features. As a result, tokens that only contain purely user-related information are designated as U-tokens, whereas tokens that contain both user and group information are categorized as G-tokens. This architectural arrangement ensures that the inputs to subsequent feature-interaction modules can be rigorously partitioned into U-tokens and G-tokens, as illustrated in Figure~\ref{fig:mixup}.

\subsection{Separation of UG Interaction}
Based on the above decomposition, the inputs to the feature-interaction module have now been separated into U-tokens and G-tokens. To ensure the reusability of U-side outputs, it is essential to prevent any G-side information from being exposed to or entangled with the U-side. In the following sections, we describe how to design an appropriate separation mask for TokenMixer \cite{zhu2025rankmixer, jiang2026tokenmixer} to guarantee the independence of U-side information.

TokenMixer architecture \cite{zhu2025rankmixer, jiang2026tokenmixer} has two core components: (1) Multi-Head Token Mixing layer, and (2) Pertoken FeedForward Network (PertokenFFN) layer:
\begin{align}
    P_{k-1} &= LN\bigl(Mixup
    (X_{k-1}) \bigr) \\
    X_{k} &= LN\bigl(PFFN(P_{k-1}) + X_{k-1}\bigr)
\end{align}

\noindent where $LN(\cdot)$ is the layer normalization function, $Mixup(\cdot)$ and
$PFFN(\cdot)$ are the multi-head TokenMixing module and the PertokenFFN module,
$ X_k \in \mathbb{R}^{T \times D} $ is the output of the $k-th$ TokenMixer
block, $T$ is the token number and $D$ is the hidden
dimension of the model. 
With UG separation, $X$ can be denote as follow:
\begin{equation}
    X = Concat(x_{u^0}, x_{u^1}, \cdots, x_{u^{n-1}}, x_{g^0}, x_{g^1}, \cdots, x_{g^{m-1}})
\end{equation}
 where $x_u$ is the U-side token, $x_g$ is the G-side token, $n$ is the number of U token, and $m$ is the number of $G-token$, where $m+n = T$. A core component of the TokenMixer is the mixup module, which splits and transposes the input tokens to enable thorough interaction across token dimensions. It first split the each token with $H$ heads:
 \begin{align}
     Split(x_{u^0})=(x_{u^0}^0, x_{u^0}^1,\cdots, x_{u^0}^{H-1})
 \end{align}
 
 Let $D'$ denote the dimension of each head, with $D' = D / H$. Accordingly, $x_{u^0}^{H-1} \in \mathbb{R}^{1 \times D'} $ for each head $h$. We then concat the splited head to get the new token $L$:

\begin{align}
        \label{eq:uginformation}
     L^h = Concat(x_{u^0}^h, x_{u^1}^h,\cdots,x_{g^{T-1}}^h), h \in [0,H-1]
 \end{align}
$L^h \in \mathbb{R}^{1 \times D'*T}$ contains both U-side and G-side information. After the above transformation, $H$ new tokens are produced. By concatenating these tokens, we obtain the output of the mixup module.
 \begin{align}
       Mixup(X) = Concat(L^0,L^1,\cdots, L^{H-1})
 \end{align}
 
At this stage, we assume that among the newly generated tokens, the first $c_u$ are U-tokens and the remaining $c_g$ are G-tokens, where $c_u + c_g = H$.  \textbf{To ensure that the computation results on the U-side can be reused, we must guarantee that the $c_u$ U-tokens has no G-side information}. To this, we first introduce a masking mechanism that removes any G-side information from these user tokens. The mask can be formula as follow:

\begin{align}
    mask_{i,j} = 
\begin{cases}
0, \quad i<c_u,\quad j>n*D', \quad  \\
1, \quad otherwise
\end{cases}
\end{align}
where $i \in [0, H-1],\quad j\in[0, T*D']$.
As shown in Fig.\ref{fig:mixup}, we can remove the G-side information from U-tokens as follows:
\begin{equation}
    X_{masked} = Mixup(X) * broadcast(mask)
\end{equation}

Thus, after the mask transformation, we obtain a new set of tokens, where the first $c_u$
tokens are designated as U-tokens and contain no G-side information. The remaining tokens correspond to G-tokens.
The other component of the TokenMixer is the PertokenFFN layer, which we decompose into two parts: a Reusable PertokenFFN and a Non-Reusable PertokenFFN. As shown in Fig.\ref{fig:mixup}, as U-side tokens are irrelevant to different candidate items in online serving, and thus these tokens are processed through the reusable PertokenFFN, while G-side tokens are passed through the non-reusable counterpart. This design ensures that the computation associated with U-side tokens only needs to be performed once for each user in serving, enabling the reduction from $\mathcal{O}(C)$ to $\mathcal{O}(1)$ for the U-token part, where $C$ is the number of candidate items in ranking stage of industrial recommenders.

\subsection{UG-Sep with Separated Residual}

When the number of U-side tokens in the input $X$, denoted by $n$, and the number of G-side tokens, denoted by $m$, are respectively equal to the numbers of U-side and G-side tokens after the Mixup operation, denoted by $c_u$ and $c_g$, a direct residual connection can be applied. As a result, U-side information remains isolated, ensuring strict preservation of U-side independence.

\begin{figure}[t]
    \centering
    \includegraphics[width=1\linewidth,clip]{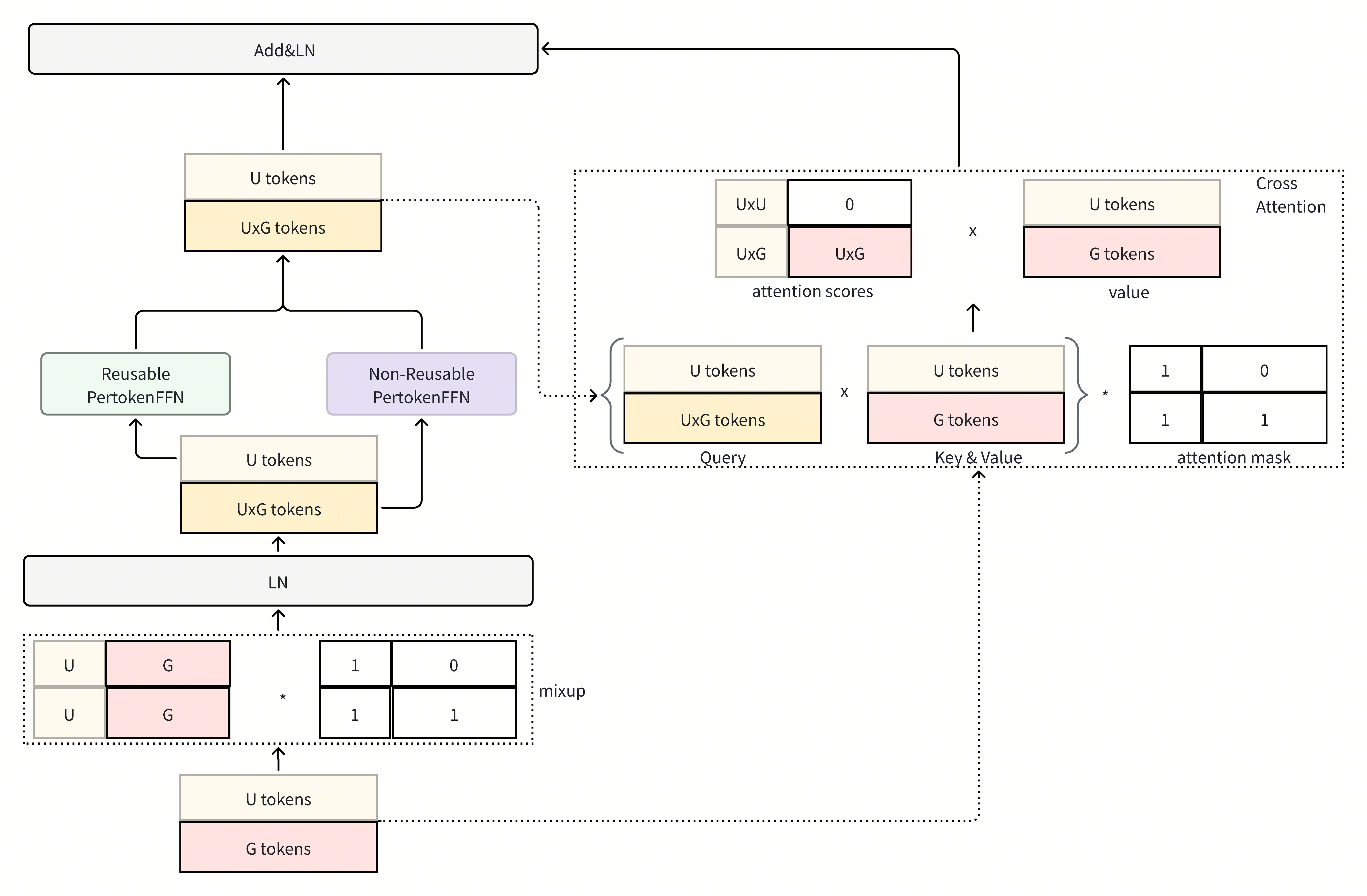}
    \caption{UG-Sep with Separated Residual}
    \label{fig:ug-resi}
\end{figure}

However, in practical applications, a pyramidal design—where the number of output tokens is smaller than that of the input—is often used to achieve higher computational efficiency, typically accompanied by residual connections. Such designs are commonly applied in deeper layers: lower layers employ the standard TokenMixer to perform full cross-token interaction, while upper layers adopt the pyramidal structure to improve parameter and computational efficiency. Consequently, when the UG-Sep mechanism is incorporated into a pyramidal TokenMixer, the number of input U-side tokens $n$ is no longer equal to the number of mixup output U-token $c_u$, which complicates the design of the residual pathway, as it must strictly prevent G-side features from being propagated into U-side tokens.

The solution is inspired by the cross-attention mechanism (see Fig.\ref{fig:ug-resi}). Specifically, the output of the Mixup followed by the PertokenFFN is used as the query, and a cross-attention operation is performed over the input tokens of the current mixer layer. The resulting output is then added back to the Mixup output as a residual connection. To maintain UG separation under this design, an additional UG mask must be incorporated into the cross-attention computation. This approach has the benefit of eliminating the constraint that the input and output tokens must maintain a fixed U/G proportion.


\subsection{Information Compensation}

\begin{figure}[h]
    \centering
    \includegraphics[width=1\linewidth]{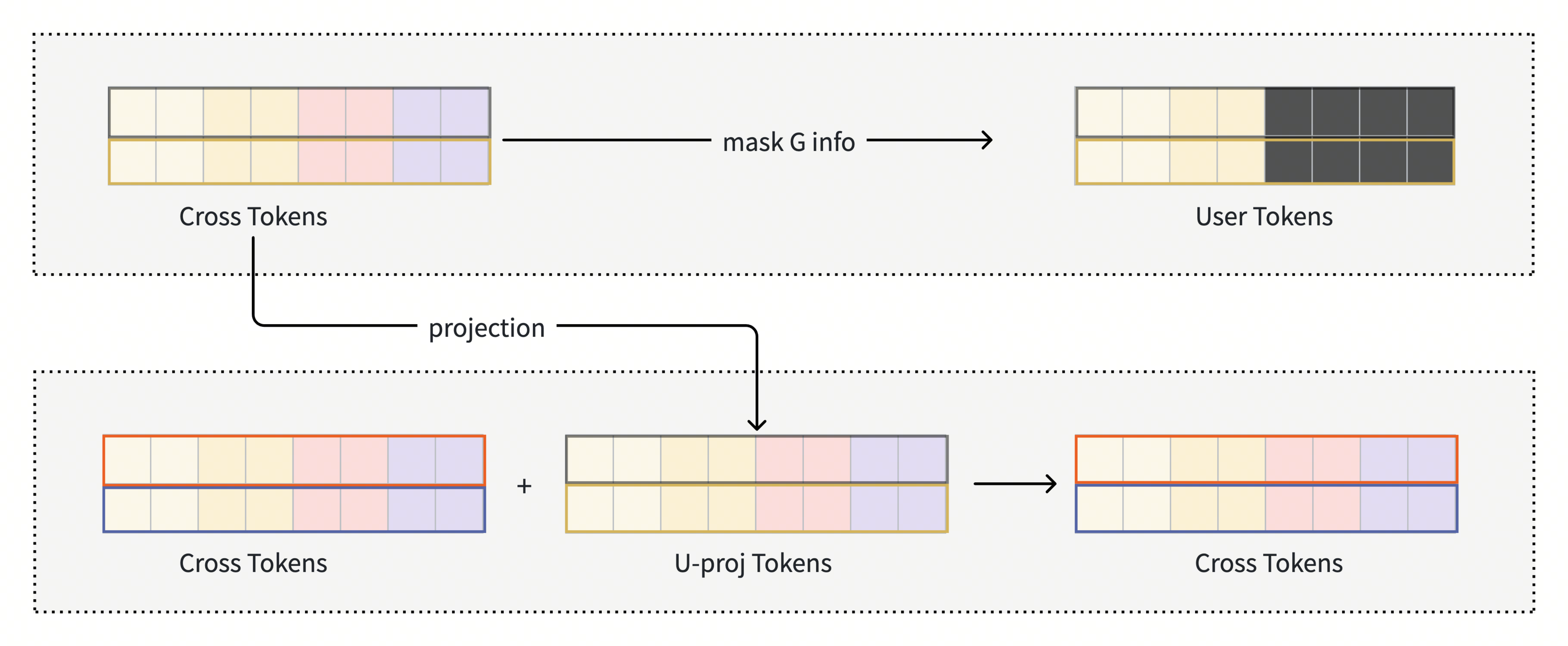}
    \caption{Information Compensation}
    \label{fig:inforation Compensation}
\end{figure}

As illustrated in Fig.\ref{fig:inforation Compensation}, the UG-Sep masking operation inevitably removes part of the U-side head information along the dimensional axes associated with G-side tokens, which introduces a certain degree of information loss. Empirically, we observe that when the proportions of U-side and G-side tokens remain close to their original allocation, this loss has a negligible impact on model performance. A plausible explanation is that the masked dimensions account for only a small fraction of the overall representation, and the residual connections are able to partially compensate for the missing information.

However, further experiments show that when the proportion of U-side tokens becomes significantly larger than that of G-side tokens (e.g., ratios of U:G become 2:1, 3:1, or even 5:1), model performance degrades substantially. In such cases, the masked G-related dimensions occupy a much larger portion of the representation space, and residual connections alone are no longer sufficient to recover the lost information.

To address this issue, we introduce an information compensation mechanism. The key idea is to explicitly inject U-side information back into the G-side representations after UG masking.
Formally, let $U \in \mathbb{R}^{c_u \times d}$ denote the U-side token representations after the Mixup operation and UG masking, $G \in \mathbb{R}^{c_g \times d}$ denote the corresponding G-token representations. We first project the U-side representations into the same feature dimension as the G-side tokens:
\begin{equation}
\hat{U} = \mathrm{Proj}(U)
\end{equation}
where $\mathrm{Proj}(\cdot)$ is a learnable linear projection.
The projected U-token are then added to the G-tokens to compensate for the masked information:

\begin{equation}
G_{\mathrm{comp}} = G + \hat{U}
\end{equation}

As shown in Fig.\ref{fig:inforation Compensation}, this operation can be interpreted as transferring the preserved U-side information back to the G-side tokens through a lightweight projection-and-addition pathway. By this, the model restores the missing contextual information caused by UG masking while still maintaining strict U-side independence, since no G-side information is injected into U-side tokens. This information compensation mechanism is particularly important in pyramidal TokenMixer settings, where the imbalance between U-side and G-side token proportions becomes more pronounced.
\begin{algorithm}[b]
\caption{In-request U-side Caching for TokenMixer}
\label{al:algo}
\begin{algorithmic}[1]
\State \textbf{Input:} candidate\_size\_tensor $\in \mathbb{R}^{M}$, \; INPUT\_U, INPUT\_G $\in \mathbb{R}^{N}$
\State \textbf{Output:} OUTPUT\_U
\Statex

\State Offset $\gets$ \textsc{Cumsum}(candidate\_size\_tensor)
\State Unique\_U $\gets$ \textsc{Gather}(INPUT\_U, Offset)
\State Unique\_U $\gets$ \textsc{TokenMixer}(Unique\_U)
\State OUTPUT\_U $\gets$ \textsc{Repeat}(Unique\_U, candidate\_size\_tensor)

\end{algorithmic}
\end{algorithm}

\subsection{W8A16 Quantizaion Method}

In the serving stage, we reuse the U-side computations. As shown in Algorithm \ref{al:algo}, for each candidate item, the cached U-side results can be directly reused, leading to a theoretical speedup proportional to the number of candidates. The expected FLOPS acceleration can be approximated by the following formulation:

\begin{align}
    ratio = \frac{c_u}{c_{u}+c_{g}}
    \label{eq: flops}
\end{align}

As shown in Eq.~\ref{eq: flops}, the FLOPs are substantially reduced by 50\%, if $c_u$ is equal to $c_g$. However, the forward latency does not decrease proportionally, as the computation is bounded by memory throughput rather than compute operations. To address this bottleneck, we propose a 8-bit weight and 16-bit activation strategy (W8A16), which significantly reduces memory traffic and leads to noticeable latency improvements.

\textbf{W8A16} is a weight-only quantization method. In memory-bound regimes, the dominant performance bottleneck arises not from arithmetic operations but from the high memory bandwidth required to load large parameter matrices. Weight-only quantization provides an effective solution by reducing the bit-width of weights while keeping activations in higher precision. By storing weights in 8-bit (FP8) format and dequantizing them on-chip during computation, the required memory traffic is reduced by up to 4× compared to FP32 and 2× compared to BF16 storage. Because activations remain in 16-bit precision (A16), the expressiveness and numerical stability of the forward pass are preserved, making the scheme compatible with large-scale recommendation models and high-throughput serving scenarios.
The computational overhead introduced by dequantization is negligible relative to the savings in memory traffic. As a result, W8A16 quantization substantially alleviates memory bottlenecks and yields pronounced latency improvements, especially in workloads with small batch sizes or models with UG Separate, where weight loading dominates execution time.

\section{Experiment}
\subsection{Experiment Settings}
\subsubsection{Datasets and Environment}
To comprehensively evaluate the effectiveness of our proposed method, we conduct offline experiments across multiple influential real-world business scenarios at ByteDance, including Douyin Feed Recommendation\footnote{https://www.douyin.com/}, Hongguo Feed Recommendation\footnote{https://novelquickapp.com/}, Chuanshanjia Ads\footnote{https://www.csjplatform.com/} and Qianchuan Ads\footnote{https://www.shoppingads.cn/}. These tasks differ substantially in terms of user scale, content modality, feedback patterns, and data distributions, offering a comprehensive testbed to evaluate the model’s generalization ability and robustness across heterogeneous environments.

All datasets are derived from real online interaction logs and user feedback signals. They contain hundreds to thousands of feature fields—including numerical, categorical, cross, and sequential features—spanning billions of user IDs and hundreds of millions of video or ad item IDs. Prior to model training, all features are transformed into sparse embedding representations to accommodate large-scale deep recommendation models. 
\begin{table}[h]
    \centering
    \caption{Evaluation of methods on different industrial datasets. U:G denotes the mixing ratio of user tokens to group tokens. “$\Delta$AUC” represents the improvement over the baseline model. Among these, an AUC decrement of 0.01-0.03\% has been verified through rigorous A/B testing to exert almost no impact on online performance. $\Delta$ Latency represents the serving latency improvement over the baseline model.}
    \label{tab:performance}
    \begin{tabular}{l|l l l l l}
        \toprule
          Scenario & U:G   & $\Delta$AUC & $\Delta$Latency \\  
        \midrule
        \multirow{4}{*}{\textbf{Douyin Feed Rec}} & base &  --& -- \\
        & 1:2 &  +0.002\%   & --  \\
         & 1:1 & -0.004\%  & -20.0\% \\
         & 3:1 & -0.013\%  & -- \\
         \midrule
         \multirow{4}{*}{\textbf{Hongguo Feed Rec}} & base &  --&-- \\
         & 1:1 & -0.018\%  & - 11.5\%\\
         & 5:3 & -0.015\%   & -- \\
         & 2:1 & -0.033\%  & --\\
         \midrule
          \multirow{3}{*}{\textbf{Chuanshanjia Ads}} & base &  --&-- \\
          & 1:1 & -0.016\%  & -12.7\%\\
         & 5:3 & -0.026\%  & --\\
         \midrule
           \multirow{2}{*}{\textbf{Qianchuan Ads}}& base & -- & -- \\
         & 1:1 & -0.024\%  & -22.0\%\\
        \bottomrule
    \end{tabular}
\end{table}

\subsubsection{Baseline}
We conduct our evaluations on the following four models: 
\begin{enumerate}
  \item \textbf{Douyin Feed Rec}:  
  We evaluate the ranking model in the Douyin recommendation system whose backbone is based on the TokenMixer\cite{zhu2025rankmixer, jiang2026tokenmixer} architecture. The training samples follow the user-level sample aggregation\cite{zhai2024actions}, allowing us to measure both training acceleration and inference acceleration.
  \item \textbf{Hongguo Feed Rec}: A short-video recommendation model used in the Hongguo platform, also built upon a TokenMixer backbone. Since its training samples are in a standard instance-level format, we only evaluate inference acceleration.
    \item \textbf{Chuanshanjia Ads}: A CTR prediction model deployed in ChuanShanJia, primarily serving off-site traffic, with a backbone composed of TokenMixer layers. Similar to Hongguo, its training samples are standard instance-level, and therefore only inference acceleration can be measured.
    \item {\textbf{Qianchuan Ads}}: An Ads CVR model in Douyin with a core architecture based on the TokenMixer.
\end{enumerate}

\subsubsection{Evaluation Metrics and Implementation Details}
For model performance evaluation, we adopt the Area Under the ROC Curve (AUC) as the primary metric. For commercial confidentiality reasons, we omit the absolute values of AUC and only retain the relative changes in AUC. Since the TokenMixer constitutes the core building block of these models, UG-Sep is primarily applied to the TokenMixer modules. During training, the same training protocol as the baseline is adopted, using large-scale industrial datasets to ensure a fair and consistent comparison.

\subsection{Offline Evaluation}

\subsubsection{AUC}

Table \ref{tab:performance} reports the AUC performance of UG-Sep under different model settings and U:G token ratios. Overall, UG-Sep maintains highly stable predictive accuracy across all evaluated models. For the Douyin ranking model, AUC variations remain within ±1e-4 across different ratios, indicating that user–group separation introduces negligible impact on model effectiveness. Similar trends are observed in the Hongguo, Chuanshanjia, and Qianchuan models, where slight AUC fluctuations are consistently below 0.0003. These results demonstrate that the proposed separation and compensation mechanisms effectively retain model expressiveness while enabling computation reuse.

\subsubsection{Different UG Token Ratios}
We also investigate the effect of different U:G token mixing ratios on model performance. As shown in Table  \ref{tab:performance}, moderate ratios such as 1:1 or 1:2 consistently achieve a favorable balance between accuracy and efficiency. In the Douyin model, increasing the proportion of user tokens leads to marginal AUC changes while improving training throughput. However, excessively skewed ratios (e.g., 3:1) do not provide additional accuracy benefits and may introduce unnecessary redundancy. Similar observations hold for Hongguo and Chuanshanjia, suggesting that UG-Sep is robust across a range of token ratios, with balanced configurations offering the most stable trade-off between accuracy and computational gains.

\begin{table}[t]
\centering
\caption{Relative training speedup of the Douyin Rec model under different U:G ratios.}
\begin{tabular}{c c c}
\hline
Model & U:G & Training Speedup (\%) \\
\hline
TokenMixer & --  & 0.0 \\
\hline
\multirow{3}{*}{TokenMixer with UG-Sep} 
& 1:2 & +5.50 \\
& 1:1 & +8.60 \\
& 3:1 & +14.8 \\
\hline
\end{tabular}
\label{tab:douyin_speedup}
\end{table}
\subsubsection{Training Acceleration}
Since the training samples in Douyin adopt the user-level training sample aggregation\cite{zhai2024actions} format, we present the resulting training throughput speedup in Table~\ref{tab:douyin_speedup}. Increasing the U-side ratio leads to stronger acceleration, with an 8.6\% speedup observed at the 1:1 setting.

\subsubsection{Serving Latency}
UG-Sep delivers consistent serving latency reductions across all evaluated models. In the Douyin production model, adopting a 1:1 UG ratio reduces serving latency corresponding to an 20\% improvement, while preserving comparable AUC. For Hongguo and Chuanshanjia, serving latency is still reduced by 11.5\% and 12.7\%, respectively. Notably, the largest Qianchuan Ads achieves the largest latency reduction of 22.0\%, highlighting that the benefits of UG-Sep become more pronounced as model size increases. Overall, these results demonstrate that UG-Sep effectively accelerates inference in all settings, and can additionally improve training efficiency when user-level aggregation is available.



\subsection{Abation Study}
To better understand the contribution of each component in our method, we perform a series of ablation studies. Specifically, we evaluate the impact of the Information Compensation module on AUC and examine how W8A16 quantization affects throughput.

\subsubsection{Information Compensation}

To further examine the role of Information Compensation, we conduct an ablation study across various U:G token ratios, as shown in Table~\ref{tab:info_comp_ablation}. Without the compensation mechanism, model performance deteriorates as the U-side proportion increases. The AUC drops from $-0.01\%$ at 1:1 to $-0.04\%$ at 2:1, and further to $-0.0\%$ at 3:1. This degradation suggests that the information obscured by UG separation masking becomes increasingly unrecoverable when the model allocates more capacity to the user-side tokens, and residual connections alone are insufficient to compensate for this loss. After enabling Information Compensation, the model exhibits notable improvements. At the 3:1 ratio, the AUC improves from $-0.06\%$ to $-0.02\%$, demonstrating a clear recovery of the masked information. Even at the highly imbalanced 5:1 ratio, the model maintains stable performance ($-0.04\%$) with compensation enabled. These findings indicate that the Information Compensation module effectively mitigates the information loss caused by UG masking and plays a crucial role in stabilizing model behavior under skewed U:G configurations. Overall, the ablation results validate that Information Compensation is a key component of our framework, especially when the model emphasizes the user-side representation.

\begin{table}[h]
\centering
\caption{Ablation study on Information Compensation under different U:G ratios.}
\begin{tabular}{lcc}
\toprule
\textbf{U:G Ratio} & \textbf{Info Compensation} & $\Delta$\textbf{AUC} \\
\midrule
1:2 & N & +0.00\% \\
1:1 & N & -0.01\% \\
2:1 & N & -0.04\% \\
3:1 & N & -0.06\% \\
\midrule
3:1 & Y & -0.02\% \\
5:1 & Y & -0.04\% \\
\bottomrule
\end{tabular}
\label{tab:info_comp_ablation}
\end{table}


\subsubsection{W8A16}

Although UG-Sep reduces FLOPs significantly, its latency improvement remains limited because the execution is dominated by memory bandwidth rather than compute operations. By applying W8A16—an 8-bit weight-only quantization scheme—the memory bound is substantially reduced, leading to significant reductions in GEMM latency across U-side tested shapes. As shown in Table \ref{tab:w8a16_ablation} , W8A16 reduces GEMM-level latency by 40.0\%–55.0\% across all evaluated configurations.

\begin{table}[h]
\centering
\caption{GEMM-level latency reduction after applying W8A16 on UG-Sep.}
\begin{tabular}{lcc}
\toprule
\textbf{BS M N K}  &\textbf{UG-Sep}& \textbf{UG-Sep + W8A16} \\
\midrule
(1,16, 1280, 2560) &-- & \textbf{-50.2\%} \\
(1,16, 1280, 640) &--& \textbf{-40.0\%} \\
(1,8, 1280, 2560) &-- & \textbf{-46.8\%} \\
(1,8, 1280, 640) &-- & \textbf{-55.0\%} \\
\bottomrule
\end{tabular}
\label{tab:w8a16_ablation}
\end{table}

\noindent Notably, W8A16 can also provide slight acceleration for the models without UG-Sep; however, the tensor shapes exhibit limited memory-bound characteristics compared to computation, resulting in limited speedups. In contrast, when model is equipped with UG-Sep, as computation cost is significantly reduced, W8A16 more effectively targets the dominant GPU access bottleneck, highlighting that weight loading—rather than computation—is the primary cost in recommendation with UG-Sep. Importantly, because activations remain in 16-bit precision, the expressiveness and numerical stability of the forward computation are preserved.


\begin{table}[h]
\centering
\caption{Online A/B Test Results on Douyin recommendation system: Key Behavioral Metrics (TokenMixer with UG-Sep vs. TokenMixer as baseline).}
\label{tab:online_ab}
\begin{tabular}{lccc}
\toprule
\textbf{Metric} &  \textbf{Change}(\%)& \textbf{$p$-value} \\
\midrule
Active Days
&-0.0020& 0.46 \\
Duration& +0.0056& 0.45 \\
Like &  -0.0511 & 0.34\\
HLT &  -0.0003 & 0.95\\
Comment &  -0.0923& 0.18 \\
\midrule
Latency&\textbf{-20.0} &--\\
\midrule
\end{tabular}
\end{table}

\begin{table}[h]
\centering
\caption{Online A/B Test Results on Chuanshanjia Ads: Key Behavioral Metrics (TokenMixer with UG-Sep vs. TokenMixer as baseline)}
\label{tab:ChuanShanJia}
\begin{tabular}{lcc}
\toprule
\textbf{Metric} & \textbf{Change} (\%) & \textbf{$p$-value} \\
\midrule
Cost            & -0.1143 & 0.45 \\
Rank Advv       & -0.1322 & 0.42 \\
Advv (Overall)  & -0.2042 & 0.32 \\
\midrule
Latency & \textbf{-12.7} &-- \\
\bottomrule
\end{tabular}
\end{table}

\subsection{Online A/B Tests}
We conducted large-scale online A/B experiments to assess the real-world impact of the proposed method by comparing TokenMixer with UG-Sep against TokenMixer on both the Douyin recommendation system and the ChuanShanJia advertising platform. As shown in Table~\ref{tab:online_ab}, key user behavioral metrics on the Douyin recommendation system—including active days, viewing duration, and interaction signals such as likes, HLT, and comments—exhibit only marginal relative changes. All observed variations are of negligible magnitude, and both positive and negative fluctuations are accompanied by consistently high $p$-values, indicating that none of the differences are statistically significant. These results demonstrate that the proposed method preserves core user engagement behaviors while achieving a substantial system-level benefit, namely a 20\% reduction in online serving latency. Similarly, the online results on the Chuanshanjia platform (Table~\ref{tab:ChuanShanJia}) show slight decreases in cost and Advv-related metrics (below 0.25\%), and all corresponding $p$-values are well above conventional significance thresholds. Therefore, these variations should be interpreted as statistically insignificant fluctuations rather than systematic performance shifts.

Overall, across both platforms, no statistically significant improvements or decrements are observed. The consistent absence of significant changes indicates that the proposed approach maintains stable online performance, preserving both user engagement and monetization efficiency.

\section{Conclusion}

This paper addresses a key efficiency bottleneck in recommender models with large parameters, where user–group feature entanglement prevents reuse of user-side computation and leads to high inference cost. We propose User–Group Separation (UG-Sep), a general framework that enables user-side computation reusable in large-scale TokenMixer architectures. UG-Sep introduces a separation and reusable PertokenFFN mechanism to disentangle user–group information flows, an information compensation strategy to preserve model expressiveness, and W8A16 quantization to alleviate memory bottlenecks exposed after reducing user-side FLOPs. Extensive offline evaluations and large-scale online A/B tests on multiple production systems at ByteDance demonstrate that UG-Sep achieves significant inference acceleration while maintaining stable accuracy and online metrics. Overall, UG-Sep provides a practical and effective solution for scaling dense recommender models in real-world industrial settings.

\balance
\bibliographystyle{ACM-Reference-Format}
\bibliography{ref.bib}

\end{document}